# Spontaneous supercrystal formation during a strain-engineered metal-insulator transition

**Authors:** O. Yu. Gorobtsov[1]*, L. Miao[2], Z. Shao[1], Y. Tan[3], N. I. Schnitzer[1], B. H. Goodge[4,5,6], J. Ruf[1], D. Weinstock[1]†, M. Cherukara[7,8], M. V. Holt[7], H. Nair[1], L.-Q. Chen[3], L. F. Kourkoutis[4,5]†, D. G. Schlom[1,5,8], K. M. Shen[2,5], A. Singer[1]

**Affiliations:**

[1] Department of Materials Science and Engineering, Cornell University, Ithaca, NY 14853, USA

[2] Department of Physics, Cornell University, Ithaca, NY 14853, USA

[3] Department of Materials Science and Engineering, Pennsylvania State University, University Park, PA 16802, USA

[4] School of Applied and Engineering Physics, Cornell University, Ithaca, NY 14853, USA

[5] Kavli Institute at Cornell for Nanoscale Science, Ithaca, NY 14853, USA

[6] Max-Planck-Institut für Chemische Physik fester Stoffe, 01187 Dresden, Germany

[7] Center for Nanoscale Materials, Argonne National Laboratory, Argonne, IL 60439, USA

[8] Advanced Photon Source, Argonne National Laboratory, Argonne, IL 60439, USA

[9] Leibniz-Institut für Kristallzüchtung, Max-Born-Straße 2, 12489, Berlin, Germany

* Corresponding author. Email: gorobtsov@cornell.edu

† Authors deceased

**Abstract:** Mott metal-insulator transitions possess electronic, magnetic, and structural degrees of freedom promising next generation energy-efficient electronics. We report a previously unknown, hierarchically ordered state during a Mott transition and demonstrate correlated switching of functional electronic properties. We elucidate in-situ formation of an intrinsic supercrystal in a $Ca_2RuO_4$ thin film. Machine learning-assisted X-ray nanodiffraction together with electron microscopy reveal multi-scale periodic domain formation at and below the film transition temperature ($T_{Film} \sim 200\text{-}250$ K) and a separate anisotropic spatial structure at and above $T_{Film}$. Local resistivity measurements imply an intrinsic coupling of the supercrystal orientation to the material's anisotropic conductivity. Our findings add an additional degree of complexity to the physical understanding of Mott transitions, opening opportunities for designing materials with tunable electronic properties.

**One-Sentence Summary:** A hierarchical phase structure, which dictates electronic anisotropy, is reported in a thin film during a metal-insulator transition.

**Main Text:**

The structural complexity of artificial nanomaterials challenges empirical understanding yet provides unique opportunities for controllable emergent functional properties. Supercrystals – structures with coexisting spatial periodicities larger than the atomic unit cell – present an intriguing opportunity to access new properties in the field of nanomaterials. Recently, metastable supercrystals were discovered to arise in an artificial $PbTiO_3/SrTiO_3$ superlattice through a nonequilibrium transition [1]. In 2D materials, moiré patterns can be interpreted as 2D supercrystals, showcasing an additional few-nanometer-large periodic electron scattering potential [2] [3]. Of particular interest for technology are potential applications in mottronics – electronics relying on Mott insulators [4]. The range of control possibilities in Mott insulators already makes them attractive for diverse applications such as memory elements [5], optical switches [6], and brain emulation [7]. In Mott insulators with a structurally driven metal-insulator transition (MIT), introducing periodic structural distortions at the nanoscale could provide a particularly potent control lever over the energy balance between competing ground states.

Calcium ruthenate, $Ca_2RuO_4$, is a Mott insulator with a structurally driven MIT between an insulating *S-Pbca* and a metallic *L-Pbca* orthorhombic phases [8] [9], where strain directly influences the transition [10] [11]. Bulk $Ca_2RuO_4$ undergoes MIT at a critical temperature of $T_{Bulk}$=357 K [12], which can be lowered by biaxial compressive strain in thin films to $T_{Film}$=230 K [13] or triggered by a surprisingly low electric field [14], making it an attractive candidate for beyond-CMOS electronic devices. Optical, x-ray, and neutron studies have suggested the presence of an intermediate nonequilibrium S'-Pbca phase arising during the current-induced MIT transition in bulk $Ca_2RuO_4$ [15] [16] [17], organized into 100-nanometer wide stripes on the surface of the crystal due to the difference in in-plane orthorhombic anisotropy between L and S' phases [15]. Neutron diffraction measurements have shown that the S and S' phases can coexist and have a different orthorhombicity (c/b-1=2% for S vs 1.3% for S' at 130 K and 10 $A/cm^2$) [16]. Further investigation has shown a possible coexistence of at least 3 phases with different *c*-axis parameters in the current-induced MIT transition [17]. The presence of nanoscale texture on the surface of bulk crystals during the transition suggests that in thin, <100 nm thick films, the coexisting phases during the transition could be made to organize on similar scales.

Here, we demonstrate that domains on the 100 nm scale can indeed be induced via cooling during the transition. Furthermore, thin-film confinement and epitaxial strain induce a stable supercrystal of multiple phases in a 34 nm thick $Ca_2RuO_4$ film (grown on $LaAlO_3$). The temperature-induced modulation of the supercrystal produces a dramatic switching of the directional electric properties. The structure and behavior of the induced supercrystal are summarized in Fig. 1. Below the MIT, an anisotropic multiperiodic structure forms with periods of 30 nm and 250 nm (Fig. 1A), correlated with anisotropy of ~10 μm scale local resistivity (Fig. 1B). As we heat the film above the MIT, the 30 nm periodicity vanishes, and a different anisotropic structure arises with a 90º in-plane rotation, coincident with a rotation of anisotropic resistivity direction by 90º (Fig. 1C).

**Superstructure in-situ evolution**

We have accessed the nanoscale heterogeneity in the $Ca_2RuO_4$ thin films by combining methods tailored explicitly to studies of lattice distortions in thin crystalline films. Namely, conventional X-ray diffraction (XRD), X-ray Bragg nanodiffraction, and cryogenic scanning transmission electron microscopy (cryo-STEM). The Bragg diffraction from the film, schematically shown in Fig. 2A depends on the local crystal structure, i.e., regions in the film with significantly different lattice periods manifest as separate Bragg peaks on the detector. We performed scanning X-ray nanodiffraction on a 5x5 μm region of the film, collecting spatial maps of the same region at different temperatures (Fig. 2A). The technique employs a raster-scan of a coherent X-ray beam focused down to 30 nm to resolve the crystal structure, structural phase, and strain distribution in thin (typically <100 nm) films [18]. The scattering volume for a 20 nm focused X-ray beam is orders of magnitude smaller than in conventional XRD, limiting the amount of signal in the satellite streak to ~1 photon per pixel per diffraction pattern (Fig. 2B). In bulk $Ca_2RuO_4$ crystals, the lattice period in the [001] direction in the low-temperature and high-temperature phases differs by ~2-3%, and XRD shows a clear difference between the Bragg peaks [12] [17]. In the strained thin film, we found an intricate diffraction pattern instead of clearly separable Bragg peaks corresponding to different phases. Apart from the circular features introduced by the focusing, the pattern averaged over all spatial positions shown in the experimental sketch (Fig. 2A) shows a strong peak in the center and a weaker diagonal streak across the detector (see fig. S5 for magnified versions of Fig. 2A). Synchrotron beamlines dedicated to nanoscale focusing of hard x-rays ensure a high degree of coherence in the illuminated spot and achieve the highest possible real-space spatial resolution, but produce complex far-field diffraction patterns that uniquely encode any crystallographic perturbations within the illuminated nano-volume into "speckled" intensity variations at the detector [19]. Although these patterns are extremely difficult to predict from the first principles and cause artifacts in purely intensity-based analysis methods, the detected complex wavefield forms an exact 1-1 relationship with its formative structure. We exploit this 1-1 relationship within the context of a scalable domain structure, which ensures that a properly sampled spatial map will produce far-field diffraction patterns with separable regions based on the re-occurrence of self-similar beam/domain interactions within the map, allowing classification of the regions without the need to either predict or deduce the local beam-sample interaction.

The complexity of the diffraction data and low signal make the manual decoupling of different features in reciprocal space challenging in a raster-scan. Therefore, we applied unsupervised machine learning to differentiate between different regions of the diffraction pattern based on their corresponding structure maps of the film [20]. As a first step, we used principal component analysis (PCA) [21] [22] [23] to understand the structure of the data and to estimate how many classes we should expect. To separate the clusters and classify the data, we used an unsupervised learning algorithm – k-means clustering (see Supplementary Materials for details; note that we used all components for classification, not the first *n* principal components). We classified regions of diffraction patterns at different temperatures separately, because the domain patterns evolve as the film goes through the transition. Especially when heating the sample fully through the transition and cooling back again, the domain pattern can change if it is intrinsic in nature (as is indeed the case, as will be demonstrated below). Fig. 2B shows the classification results for diffraction patterns measured on the heating branch at a fixed temperature of 250 K, slightly above the transition temperature ($T_{Film}$) of the $Ca_2RuO_4$ film and below the transition temperature of the bulk $Ca_2RuO_4$ ($T_{Bulk}$). The colors and numbers mark the clusters of pixels on the detector with correlated patterns in real space (sample maps). The translation of class information onto the diffraction pattern (Fig. 2B) yields intuitive and discrete regions. Clusters 1 and 4 are positioned along the diagonal and off-center. These clusters are associated with satellite peaks visible in XRD and identify the structure that emerges at low temperatures. Clusters 2 and 5 encompass areas of the highest relative intensity in the center of the diffraction pattern, likely connected to a high-temperature structure as the center peak is present at 300 K (see fig. S7). Cluster 3 represents a diffuse halo around the center peak. Similar data has been linked to diffuse scattering emerging from localized strain regions due to crystal defects [20], also present at 300 K.

Maps of the corresponding structural phase distribution (Fig. 2C) show domains of the structural phases in the film. A subset of these maps, corresponding to the clusters 1 and 4 as well as cluster 2, exhibit characteristic diagonal stripes oriented along the orthorhombic $[010]_o$ crystallographic direction. The typical size of the domains is on the order of hundreds of nm. Notably, an apparent degree of anticorrelation between the spatial maps of clusters 1+4 and 2 shows domain separation in real space, similar to other mott insulators [24]. Unexpectedly, the spatial map of cluster 5 presents an alternative stripe orientation, oriented along $[100]_o$ direction, that is, orthogonal to phase distribution domains. The data collected at 250 K demonstrates that the system hosts three distinct structures near the MIT temperature (Fig. 2C): A – a nanotextured phase with a distribution shown in Fig. 2C (clusters 1 and 4), B – an anticorrelated nanotextured phase with a distribution shown in Fig. 2C (cluster 2), and C – a third structure shown in Fig. 2C (cluster 5) with diagonal stripes oriented orthogonally to the stripes observed in structure (A) and (B).

We performed in-situ X-ray nanodiffraction at different temperatures through the phase transition to elucidate the development and nature of the strain modulation morphology. Figure 3 shows maps for clusters 1 and 4 for different temperatures on a heating and cooling branch. After a heating and cooling cycle, the film does not reform in the same configuration, meaning that the domain formation mechanism is intrinsic and the domain pinning by structural defects is weak. Fourier transform of a spatial map projection at 45 degrees (along the diagonal) shows a periodicity of 250 nm present in the structure. Note that the static domain size is similar to that observed on the surface of bulk crystals in experiments with applied current [15], hinting at an intrinsic length scale in the material present in bulk and thin films. The domains at lower temperatures have a stripe shape with a preferential orientation; however, on the cooling branch, it is visible in the 240 K domain map that the domains start as "pools" and only later become stripe-shaped. It has previously been observed in ruthenates under external stress that the new structural phase domains tend to grow maximally in the direction of the maximum stress [25], and the striped structure on the surface of bulk $Ca_2RuO_4$ in [15] has also been connected to the stress direction. In the strained film that we measured, the direction of the maximal stress coming from the substrate is the $[010]_o$ direction [12], which would be oriented at 45º to the edges of the measured square region.

At 300 K, the clustering reveals that the "medium-scale" (MS) structure visible in cluster 5 at 250 K persists despite the absence of a satellite streak associated with the low temperature phase domains. The MS structure is not visible at temperatures below 200 K. In short, the low temperature 250 nm stripe structure is only present at or below $T_{Film}$, while a perpendicular MS stripe structure is only present at or above $T_{Film}$.

To test if the supercrystal pattern has an impact on electric transport we measured the local resistivity anisotropy through the MIT transition by using microscale devices (see Fig. 1B, figs. S2-S4). When the resistivity is measured along two orthogonal crystallographic directions $[100]_o$ and $[010]_o$, the conductivity is higher along $[010]_o$ by a factor of about 3 at room temperature. The direction of the lowest resistivity changes from $[010]_o$ by 90º to $[100]_o$ when cooled through the MIT. We conclude that at the nanoscale, the structural and electronic inhomogeneities are strongly correlated.

**Fine structure characterization, verification and theory**

To better characterize the 30 nm periodic structure, we collected synchrotron XRD measurements on a 35 nm thick $Ca_2RuO_4$ film with a 500 μm large parallel beam. The data demonstrate a Bragg peak shape markedly different from bulk observations (fig. S5; similar to a 17 nm film in Ref. [13]). At a temperature of T=300K ($T_{Film}<T<T_{Bulk}$), the diffraction pattern does not consist of a single Bragg peak. Rather it is twinned due to a tens-of-microns-sized domain structure imposed on the film by the LaAlO$_3$ [26]. Below T$_{Film}$ additional satellite peaks around each twin-peak are present. The observed diffraction pattern is reminiscent of the Bragg peaks typically found in ferroelectric thin films caused by a herringbone twin domain structure (see, for example, Fig. 9C in [27] or Fig. 3 in [28]). The satellite peak orientation that we observed suggests a herringbone pattern oriented out of the film plane. The distance between the fringes of the satellite streaks indicates the average nanotexture domain size of ~30 nm projected on the film surface (larger than in a thinner film reported in [13]). To directly investigate the atomic structure associated with herringbone-like domain pattern at low temperatures, we have performed temperature-dependent STEM measurements (see *Supplementary Materials* for details) finding a herringbone-like domain structure when the film is viewed in cross section along the zone axis at cryogenic temperatures below T$_{film}$, and a homogeneous structure at room temperature (Fig. 4A).

To further explain the structure formed during the transition, a computational approach was adopted. Following Khachaturyan's microelasticity theory [29], we investigated the favorable interface directions between *S-Pbca* and *L-Pbca* phases, which minimizes the elastic free energy. Given the lattice parameters for the *S-Pbca* phase ($a_S$ = 5.60 Å, $b_S$ = 5.39 Å, $c_S$ = 11.76 Å) and the *L-Pbca* phase ($a_L$ = 5.36 Å, $b_L$ = 5.35 Å, $c_L$ = 12.26 Å), the favored interface normal directions will be close to $[021]_o$ and $[02\bar{1}]_o$, with a 48° tilting angle measured from the $\{001\}_o$ plane as shown in Fig. 4b, which is consistent with the diffraction patterns and the STEM images. To further elucidate the pattern observed on the cross-sectional image of the film, we used the phase-field method to simulate the *L-Pbca/S-Pbca* structure [30] [31]. With a -1.6% epitaxial strain applied on a 34 nm-thick film cross section (taking the *L-Pbca* phase as the reference), the result of the temporal relaxation from a random initial structure is shown in Fig. 4C. In the plot, an order parameter $u$ is used to distinguish between the phases. The plot shows tilted stripes in two directions formed with the coexistence of L/S phases, with a ~30 nm periodicity, which agrees with the experimental observations. The tilting angle of the stripes are close to the microelasticity predictions, which suggests that the formation of the nanostructures is a result of minimization of the elastic free energy. We have also investigated the change in the lattice parameter along $c$ axis of the orthorhombic structure, by computing $c = c_L(1 + \varepsilon_{33})$, where $\varepsilon_{33}$ is the out-of-plane normal strain component. The results depicted in Fig. 4D show a distortion along the $c$ axis, which corresponds to the change in interplanar distances as observed using HAADF-STEM and the diffraction streak in the nanodiffraction data.

**Concluding remarks**

We have found a stable, oriented supercrystal phase in a strained Ca$_2$RuO$_4$ film. The comparative (perpendicular) orientation of the streaks in the x-ray diffraction data produced by the fine 30 nm structure to the 250 nm domain structure can be explained by a self-consistent scheme of the total structure below T$_{film}$ presented in Fig. 1A, though we cannot state a uniqueness of this solution, as is further discussed below. Below T$_{film}$, the stripes of the S-Pbca and nanotextured (30 nm periodic) phases are oriented along the $[010]_o$ direction. At high temperatures, a different structural morphology is found, with stripes of a different crystal lattice period and unclear origin oriented along $[100]_o$ – perpendicular to the low-temperature structure. We hypothesize that the stripes at room temperature appear due to elastic frustration – the temperature is lower than the bulk MIT temperature – producing a strained superstructure. The two structural morphologies coexist around the film MIT temperature of ~250 K. During cooling, the disappearance of high-temperature morphology and appearance of low-temperature supercrystal results in an apparent rotation of the stripes. We have found that the direction of the preferential local electronic transport correlates to the orientation of the stripes comprising the ~250 nm periodicity order of the supercrystal. The correlation between structural and conductive properties suggests that the stripes of the conducting phase are responsible for the variation in the transport direction. Dynamical mean-field theory indicates that in $Ca_2RuO_4$, the Mott transition is locally driven by the change in the structure [9]. The coincident rotation of the structural stripes and the easy axis in anisotropic transport, as presented here, suggests that the structure of the supercrystal also governs the larger-scale direction of electron flow.

Our findings are particularly significant in the context of multiple recent studies of moiré patterns/superlattices that achieved previously unreachable electronic, photonic, magnetic and topological functionalities in 2D materials [3] [32] [33] [34] [35]. Similarly, induced supercrystals could potentially expand the palette of achievable properties in thin films. Our research shows how strain can be employed to manipulate supercrystals in thin films. Furthermore, the existence of metastable superlattices photoinduced in heterostructures [1] suggests a potential complex ultrafast behavior of the supercrystal, as well as the potential for further tuning the physical properties of the supercrystal phase by combining multiple layers of $Ca_2RuO_4$ films in a heterostructure.

While our data leaves some structural ambiguities, future advancements in spatially resolved electronic measurements, x-ray methods, and machine learning might provide further clarity. From our data, we cannot distinguish between the two possible orthogonal relative orientations of electronic transport and supercrystal order or resolve electronic transport on a nanometer length scale. This question might be further clarified by spatially resolved electronic measurements, e.g., SNOM [15]. The exact distribution of the fine 30 nm structure within the larger 250 nm structure within the film cannot be resolved with currently available diffraction or transmission methods. This allows multiple possible constructions of a supercrystal from available data similar to the one presented in Fig. 1A. With advances in x-ray sources [36], nanodiffraction will likely be able to resolve fine structure directly within the larger domains. The herringbone-like phase distribution significantly expands the surface area, raising questions about interface physics. X-ray spectroscopy at a nanoscale would also be available on the new sources. Machine learning approaches for multiscale hierarchical structures will be useful to further identify structures with better spatial and especially temporal resolution, using e. g. additional limitations on input [37].

The new hierarchies of periodicity presented here highlight prospects for inducing novel properties in nanomaterials. In particular, the impact of the superstructure on the preferential transport direction offers compelling opportunities for controlling electronic properties. The switchable behavior of the transport direction, modifiable by temperature changes, is especially intriguing for electronic/mottronic applications. Our work showcases a rotation of transport anisotropy in response to temperature modulation. Another potential avenue for investigation would be the possibility of rotation of structural order by the application of current [38] or light. Our initial attempts to rotate the direction of the easy resistivity axis by current were unsuccessful, suggesting a large energy barrier for stripe rotation. The supercrystal hierarchies and their potential effects on nanomaterial physical properties, especially in electronic applications, are promising areas for further exploration.

**Acknowledgments:** We thank Sergey Antopolsky for useful discussions regarding machine learning algorithms. The work was primarily supported by U.S. Department of Energy, Office of Science, Office of Basic Energy Sciences, under Contract No. DE-SC0019414 (O.G., Z.S., D.G.S., K.S., A.S., H.N.). This research was funded in part by the Gordon and Betty Moore Foundation's EPiQS Initiative through grant nos. GBMF3850 and GBMF9073 to Cornell University. Work performed at the Center for Nanoscale Materials and Advanced Photon Source, both U.S. Department of Energy Office of Science User Facilities, was supported by the U.S. DOE, Office of Basic Energy Sciences, under Contract No. DE-AC02-06CH11357. Device fabrication were, in part, facilitated by the Cornell NanoScale Facility, a member of the National Nanotechnology Coordinated Infrastructure (NNCI), which is supported by NSF Grant No. NNCI-2025233. Transmission electron microscopy work was supported by the National Science Foundation (Platform for the Accelerated Realization, Analysis, and Discovery of Interface Materials (PARADIM)) under Cooperative Agreement No. DMR-2039380 and made use of the Cornell Center for Materials Research Shared Facilities, which are supported through the NSF MRSEC program (DMR-1719875). N.S. was supported by the NSF Graduate Research Fellowship (DGE-2139899). B.H.G. was supported by PARADIM (NSF DMR-2039380). L.F.K. acknowledges support by the Packard Foundation. Y.T and L.Q.C's work is supported by the Penn State Center for Nanoscale Sciences, an NSF MRSEC under the grant number DMR-2011839 (2020 - 2026). C.D.'s effort is supported by the U.S. Department of Energy, Office of Science, Office of Basic Energy Sciences, under Award Number DE-SC-0012375. V.A.S., J.W.F., and L.Q.C. acknowledge the U.S. Department of Energy, Office of Science, Office of Basic Energy Sciences, under Award Number DE-SC-0012375 for support studying complex-oxide heterostructure with X-ray scattering.

**Funding:**

    U.S. Department of Energy, Office of Science, Office of Basic Energy Sciences, under Contract No. DE-SC0019414

    Gordon and Betty Moore Foundation's EPiQS Initiative through grant nos. GBMF3850 and GBMF9073

    DOE's Office of Science under contract DE-AC02-06CH11357

    NSF DMR-2104427

**Author contributions:**

Conceptualization: O.G., A.S.

Methodology: O.G.

Investigation: O.G., L.M., Z.S., N.S., D.W., Y.T., B. G.

Visualization: O.G., L.M., Z.S., Y.T., N.S.

Software: O.G., Z.S.

Resources: M. C., M. H., L. M., H. N., J. R.

Funding acquisition: A.S., D.G.S., K.S., L.-Q.C., L.K.

Project administration: O.G., A.S.

Supervision: O.G., A.S., D.G.S., K.S., L.-Q.C., L. K.

Writing – original draft: O.G.

Writing – review & editing: All authors have participated in review and editing.

**Competing interests:** Authors declare that they have no competing interests.

**Data and materials availability:** The cryogenic scanning transmission electron microscopy data of the Ca2RuO4 thin film are available for download at the Platform for the Accelerated Realization, Analysis, and Discovery of Interface Materials database (PARADIM) Underlying nanodiffraction data are available upon request.

**Supplementary Materials**

Materials and Methods

Figs. S1 to S9

**Figures**

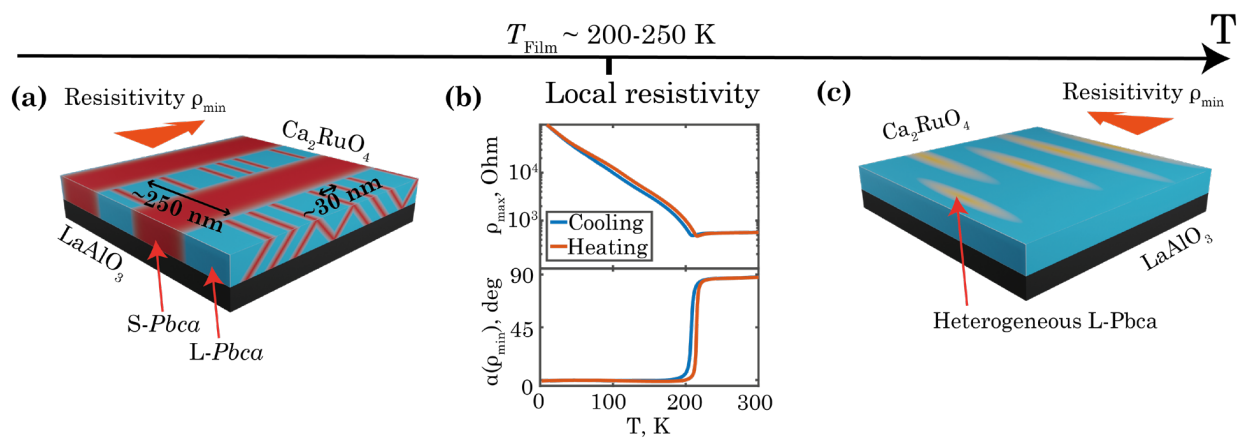

**Fig. 1. Schematic supercrystal structure in $Ca_2RuO_4$.** **(A)** Schematic representation consistent with x-ray data of the supercrystal structure below MIT. **(B)** Local resistivity of the film (top) and orientation of the easy axis in resistivity (bottom) as a function of temperature. **(C)** Schematic representation of the anisotropic structure above MIT.

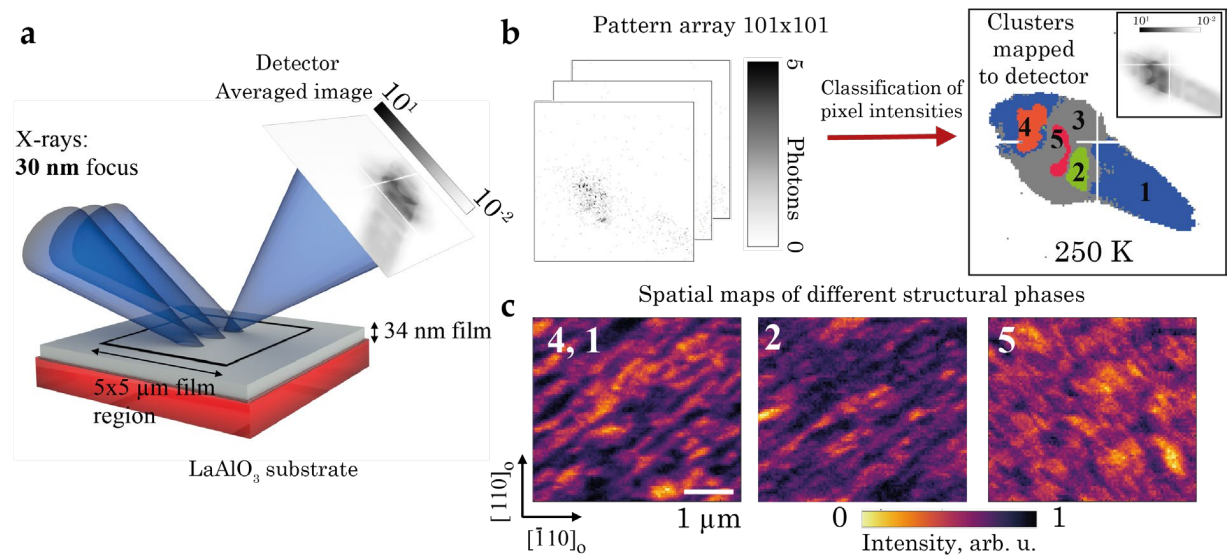

**Fig. 2. X-ray nanodiffraction and classification by machine learning.** **(A)** Schematic of a nanodiffraction experiment; **(B)** Nanodiffraction patterns measured at 250 K are classified with k-means algorithm, with regions on the diffraction pattern marked according to classification results (inset – average over diffraction pattern); **(C)** Real space maps corresponding to different classes.

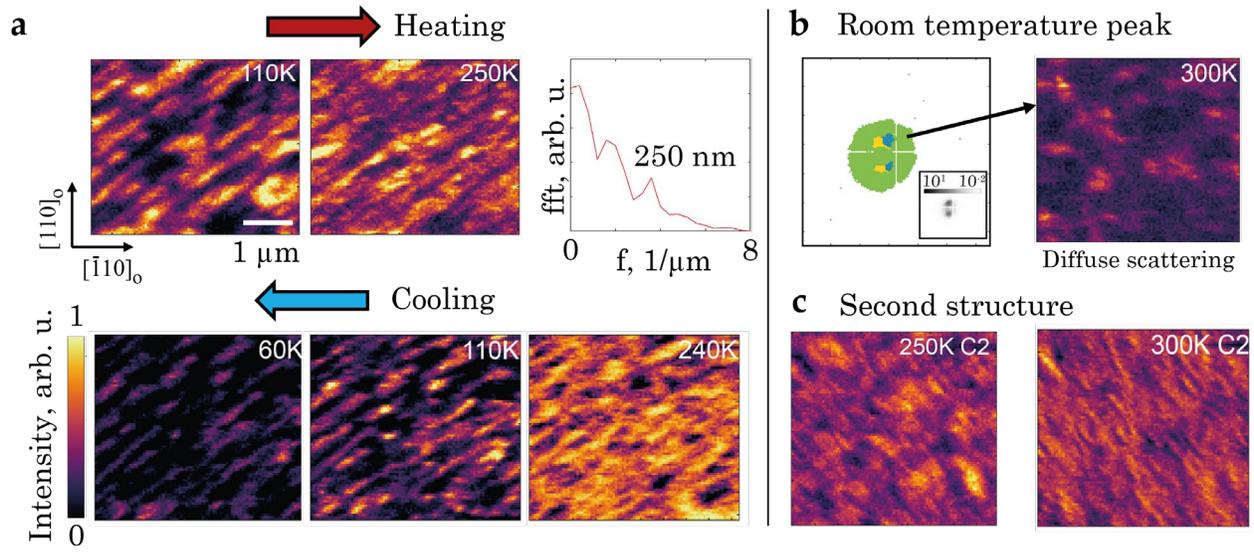

**Fig. 3. Phase distribution as a function of temperature. (A)** Maps of the phase distribution in the same region of the film at different temperatures. Structure corresponding to clusters 4,1 in Fig. 2. **(B)** Regions of the room temperature (300 K) diffraction pattern marked according to classification results for that temperature (inset – average over diffraction patterns) and defects visible in the diffuse scattering [20]. **(C)** Maps of the phase distribution of the second structure corresponding to "yellow" (left) cluster in **(B)** at room temperature and cluster (5) in Fig. 2.

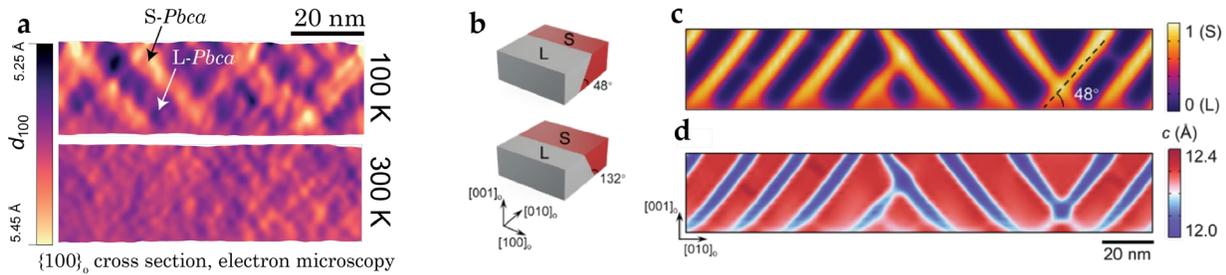

**Fig. 4. Electron microscopy and mechanical model. (A)** Real space maps of the (010) inter-planar spacing ($d_{010}$) extracted from STEM images of a cross-section of the film on the [100] zone axis (unprocessed images in fig. S1). **(B)** Prediction of the favored interface directions according to microelasticity theory. **(C)** Distribution of order parameter $u$ in the $\{100\}_o$ cross section. $u = 0$ stands for *S-Pbca* phase and $u = 1$ for *L-Pbca* phase. **(D)** The map of lattice parameter $c$ in the cross section obtained by $c = c_L(1 + \varepsilon_{33})$.

# Supplementary Materials for

## Spontaneous supercrystal formation during a strain-engineered metal-insulator transition

**Authors:** O. Yu. Gorobtsov[1]*, L. Miao[2], Z. Shao[1], Yu. Tan[3], N. I. Schnitzer[1], B. H. Goodge[4,5,6], J. Ruf[1], D. Weinstock[1]†, M. Cherukara[7], M. Holt[7], H. Nair[1], L.-Q. Chen[3], L. F. Kourkoutis[4,5]†, D. G. Schlom[1,5,8], K. M. Shen[2,5], A. Singer[1]

**Affiliations:**

[1] Department of Materials Science and Engineering, Cornell University, Ithaca, NY 14853, USA

[2] Department of Physics, Cornell University, Ithaca, NY 14853, USA

[3] Department of Materials Science and Engineering, Pennsylvania State University, University Park, PA 16802, USA

[4] School of Applied and Engineering Physics, Cornell University, Ithaca, NY 14853, USA

[5] Kavli Institute at Cornell for Nanoscale Science, Ithaca, NY 14853, USA

[6] Max-Planck-Institut für Chemische Physik fester Stoffe, 01187 Dresden, Germany

[7] Argonne National Laboratory, Argonne, IL 60439, USA

[8] Leibniz-Institut für Kristallzüchtung, Max-Born-Straße 2, 12489, Berlin, Germany

\* Corresponding author Email: gorobtsov@cornell.edu

† Authors deceased

**The PDF file includes:**

Materials and Methods
Figs. S1 to S#

## Materials and Methods

Sample Preparation

A Ca2RuO4 thin film approximately 34 nm thick was grown in a Veeco Gen10 molecular-beam epitaxy (MBE) system on a (001)$_{pc}$-oriented LaAlO$_3$ substrate from CrysTec GmbH, where pc denotes pseudocubic indices. The film was grown at a substrate temperature of 870 °C as measured using a pyrometer operating at 1550 nm. Elemental calcium (99.99% purity) and elemental ruthenium (99.99% purity) were evaporated from a low-temperature effusion cell and an electron beam evaporator, respectively. The films were grown with a calcium flux of $1.8 \times 10^{13}$ atoms·cm$^{-2}$s$^{-1}$ and a ruthenium flux of $1.7 \times 10^{13}$ atoms·cm$^{-2}$s$^{-1}$ in a background pressure of $7 \times 10^{-7}$ Torr of ozone (10% O3 + 90% O2). At the end of the growth, the shutters on both the calcium and ruthenium sources were closed, and the sample was cooled down to 250 °C in the same background pressure of ozone as used during the growth. All data presented in this paper were collected on the same film.

Electron Microscopy

Scanning transmission electron microscopy (STEM) measurements were performed on a cross-sectional lamella prepared via focused ion beam (FIB) lift-out with a Thermo Fisher Helios G4 UX FIB. Cryogenic and room temperature high-angle annular dark-field (HAADF) STEM imaging were performed sequentially on the same day on a FEI/Thermo Fisher Titan Themis 300 CryoS/TEM with a Gatan 636 double tilt liquid nitrogen cooling holder operating at 300 kV with a 30 mrad probe convergence semi-angle. For high-precision structural measurements, series of 20 rapid-frame images (~1.6 sec. per frame) were acquired, aligned, and averaged by a method of rigid registration optimized to prevent lattice hops [38] to recover high signal-to-noise ratio, high fidelity atomic resolution images. The $d_{100}$ interplanar spacing was extracted from the HAADF-STEM images using a strain mapping technique developed by Smeaton et al. [39]. Unprocessed images and corresponding interplanar spacing maps are shown in fig. S1.

Device Fabrication and local resistivity tensor measurements

To measure the local in-plane resistivity tensor (over a 6-μm-wide region) of the (001)-oriented Ca$_2$RuO$_4$ thin film, regions of the sample were fabricated into Hall bars with a channel width of 6 μm along various directions as defined by photolithography, and Ar$^+$ ion milling. Pt/Ti pads were sputtered on the thin films to form Ohmic contacts. The resistivity tensor elements $\rho_{xx}$, $\rho_{yy}$, and $\rho_{xy}$ are obtained from the longitudinal and transverse resistances of the Hall bars, which are measured in a Physical Property Measurement System (PPMS, Quantum Design, Inc.) The resistivity tensor $\rho_{ij}(T)$ is then diagonalized at each temperature to obtain two eigenvalues, a larger $\rho_{max}(T)$ and a smaller $\rho_{min}(T)$, as well as the angle $\alpha[\rho_{min}]$ between the eigenvector of $\rho_{min}(T)$ and the pseudocubic $[110]_{LAO}$ directions. A phase shift of $\alpha(T)$ by 90º is observed when the temperature of the sample crosses the metal-to-insulator transition.

X-ray measurements

All scanning X-ray nano-diffraction experiments were conducted at the 26 ID Hard X-ray Nanoprobe Beamline operated by the Center for Nanoscale Materials at the Advanced Photon Source, Argonne National Laboratory. Liquid helium cryostat was used to cool the samples. A liquid-nitrogen cooled Si (111) double crystal monochromator (DCM) gave energy resolution $\Delta E/E = 1.7 \times 10^{-4}$ and the x-ray photon energy of 10 keV. A Fresnel zone plate with an outside diameter of 150 μm and outermost zone width of 16 nm combined with an order sorting aperture was used to focus the collimated x-ray beam to 30 nm (FWHM) diameter with the flux of $\sim 10^9$ photons/s. In the horizontal scanning direction, the projection of the beam onto the sample is 30 nm / sin(θ), where θ is the angle of the incident beam (about 23° for the Ca$_2$RuO$_4$ 002 reflection at 10 keV). The horizontal footprint of the beam was, therefore, ~80 nm. We collected raster scans within a single LaAlO$_3$ domain, which are on the order of 50 μm.

Machine learning

Machine learning section consists of two parts: principal component analysis (PCA) for visualization and verification of the clusterization (see also fig. S9) and the clustering itself (see also fig. S8). The number of observations is the number of pixels (binned by 2 in every direction), and the number of variables is the number of probing points in the map. We used k-means clustering on the full dataset for every temperature, not on the first few principal components produced by PCA, due to a relatively low amount of variance explained by the first components (first 3 components only explain approximately 10%). Such a low proportion is due to the high noisiness of the data due to low signal (1-2 photons average per pixel per pattern). Note that clustering has to be performed for every temperature separately, due to the changing contrast of the domain pattern in real space and the intrinsic nature of the domain pattern (making them uncorrelated between repeated transitions).

**Supplementary Text**

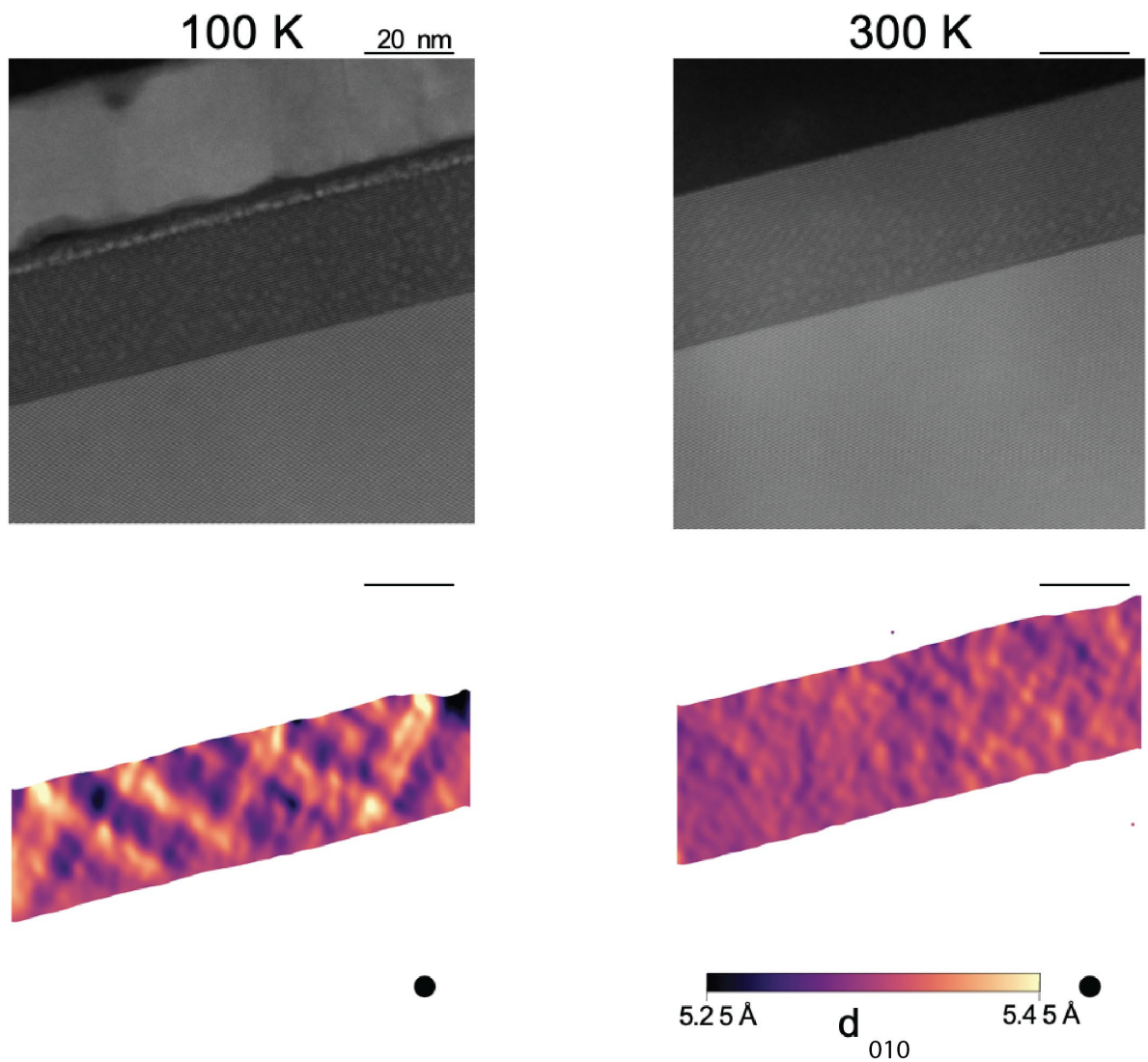

**Fig. S1.**
Full STEM datasets and corresponding lattice parameter maps. Real space maps of the (010) inter-planar spacing ($d_{010}$) extracted from STEM images of a cross-section of the film on the [100] zone axis.

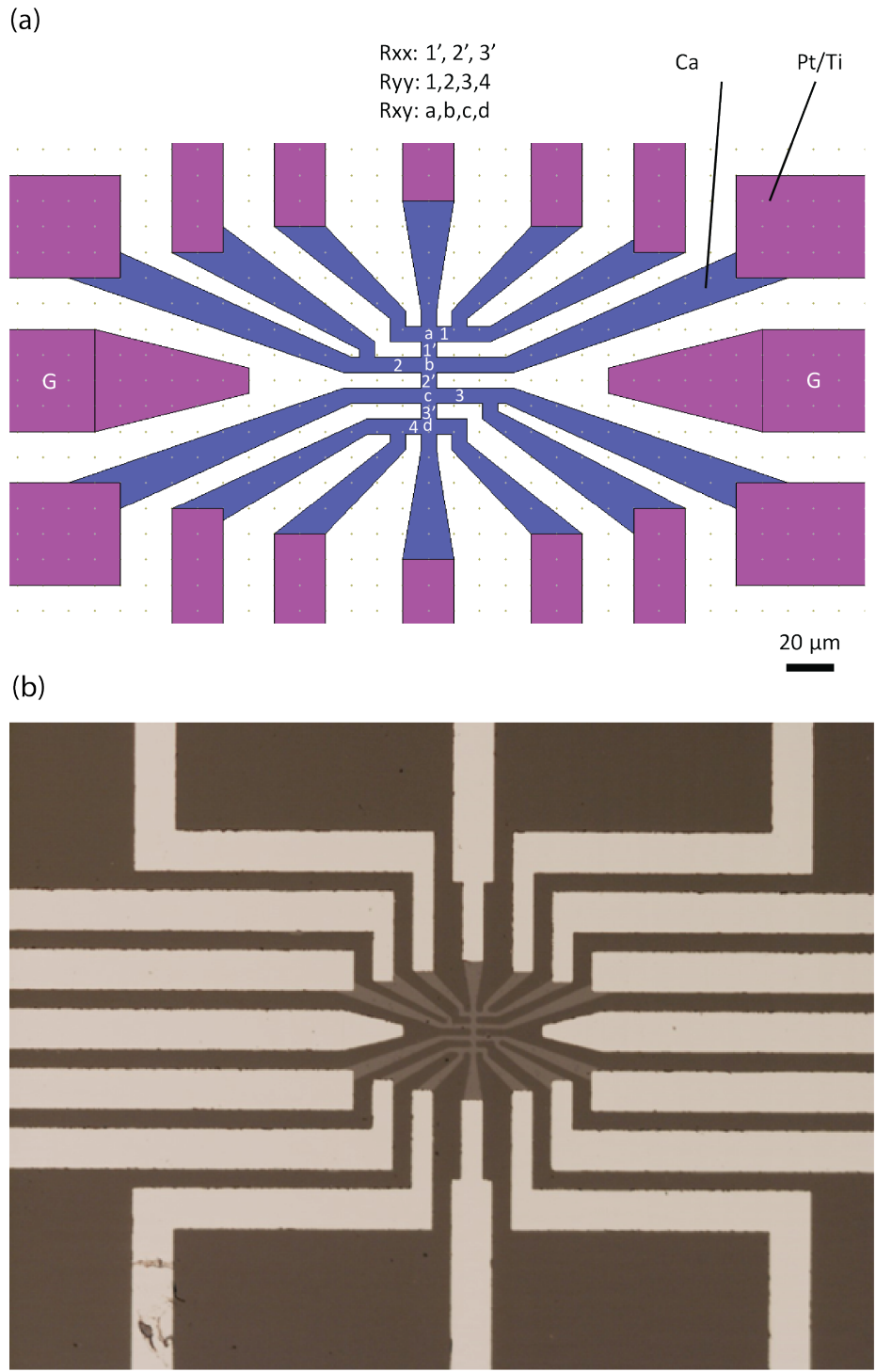

**Fig. S2.**
A. A schematic of the patterned film and contact metallization used to measure the in-plane resistivity tensor of the (001)-oriented $Ca_2RuO_4$ film. B. Photo of a typical patterned film for the resistivity tensor measurement.

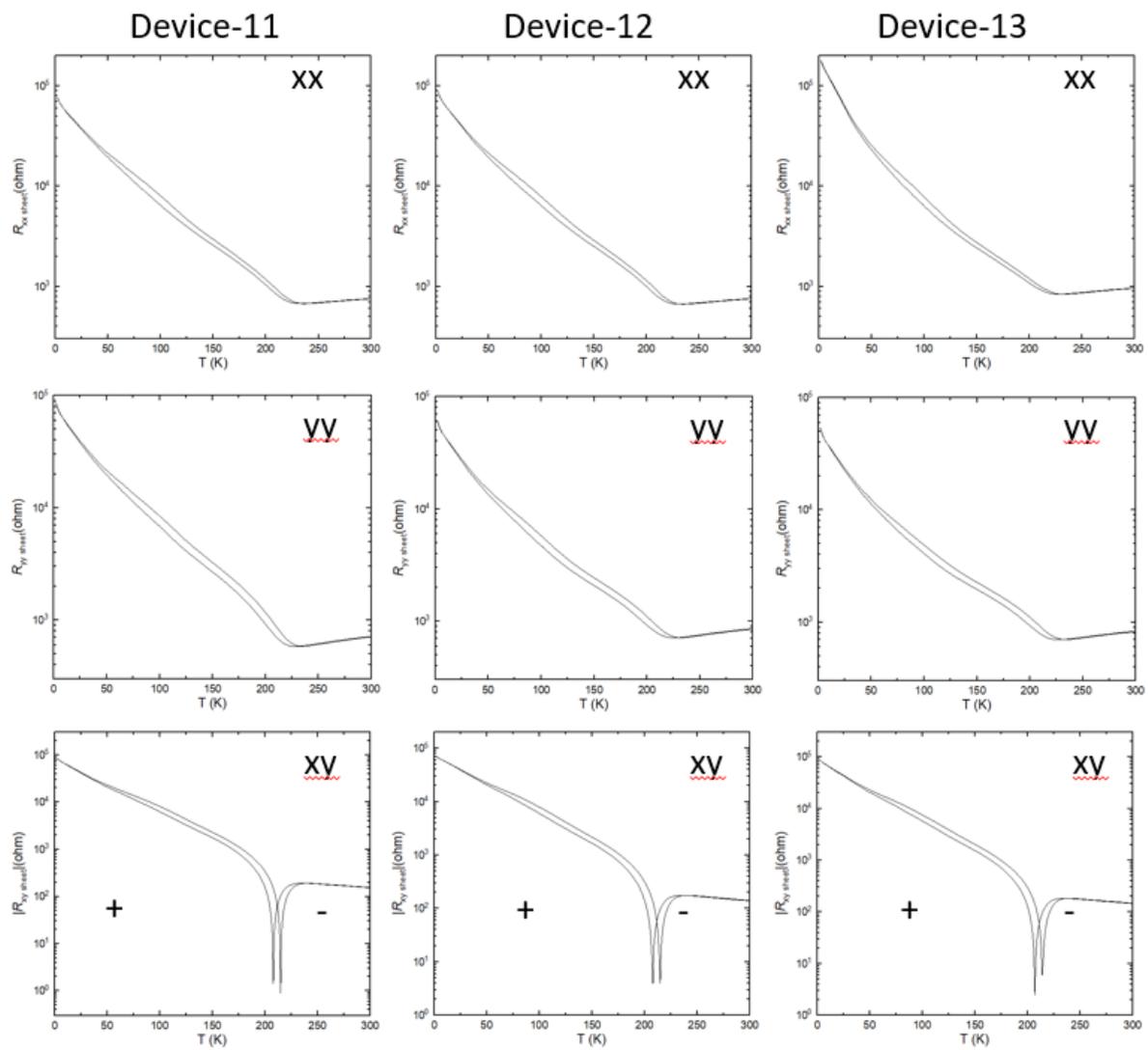

**Fig. S3.**
Individual resistivities for different directions for device 1.

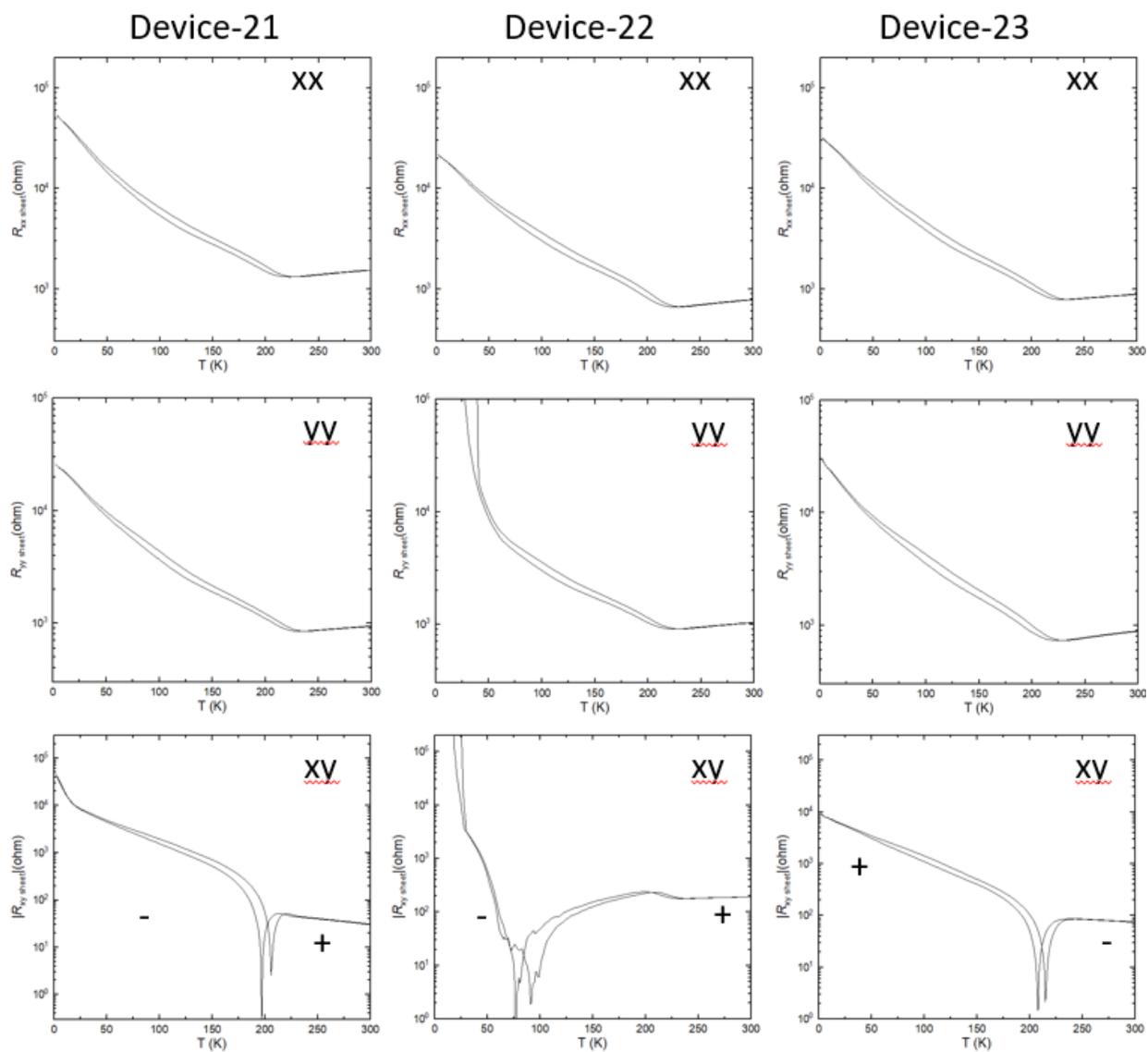

**Fig. S4.**
Individual resistivities for different directions for device 2.

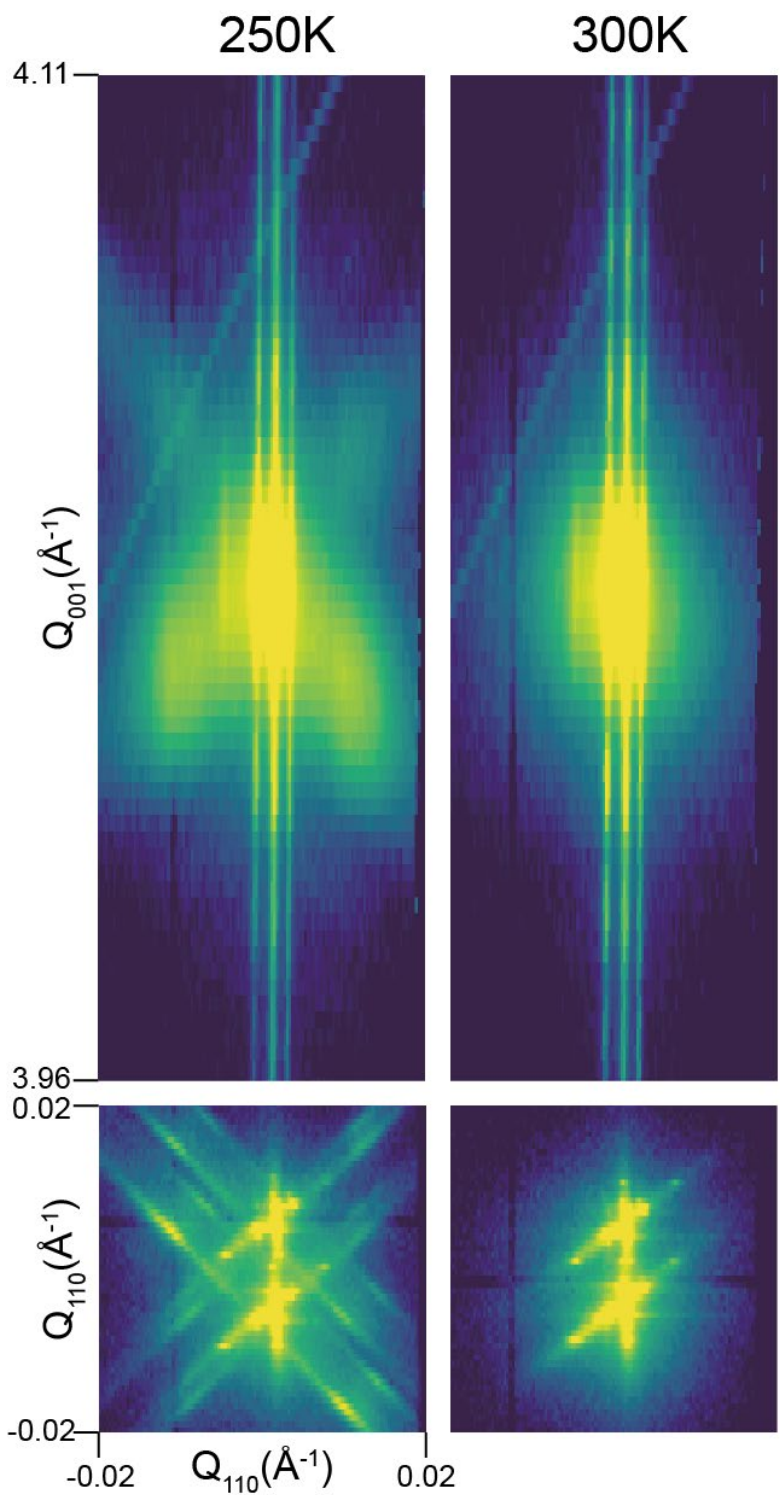

**Fig. S5.**

Examples of XRD diffraction patterns taken with a 500 μm diameter parallel incident beam at different temperatures in different projections.

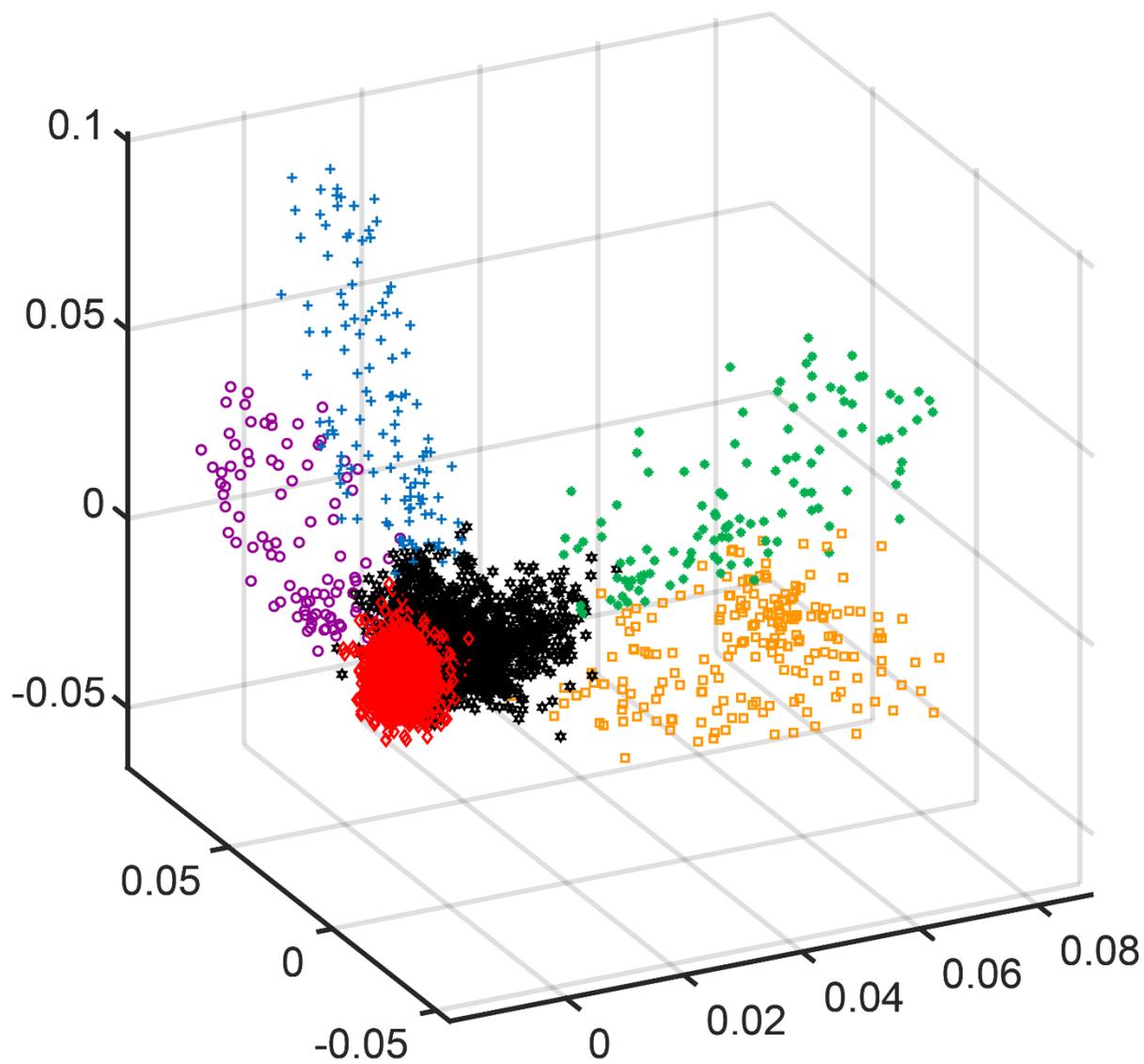

**Fig. S6.**
Example of cluster separation of detector pixels marked in 3 principal components (110 K, heating). Different clusters are marked by different color and marker type.

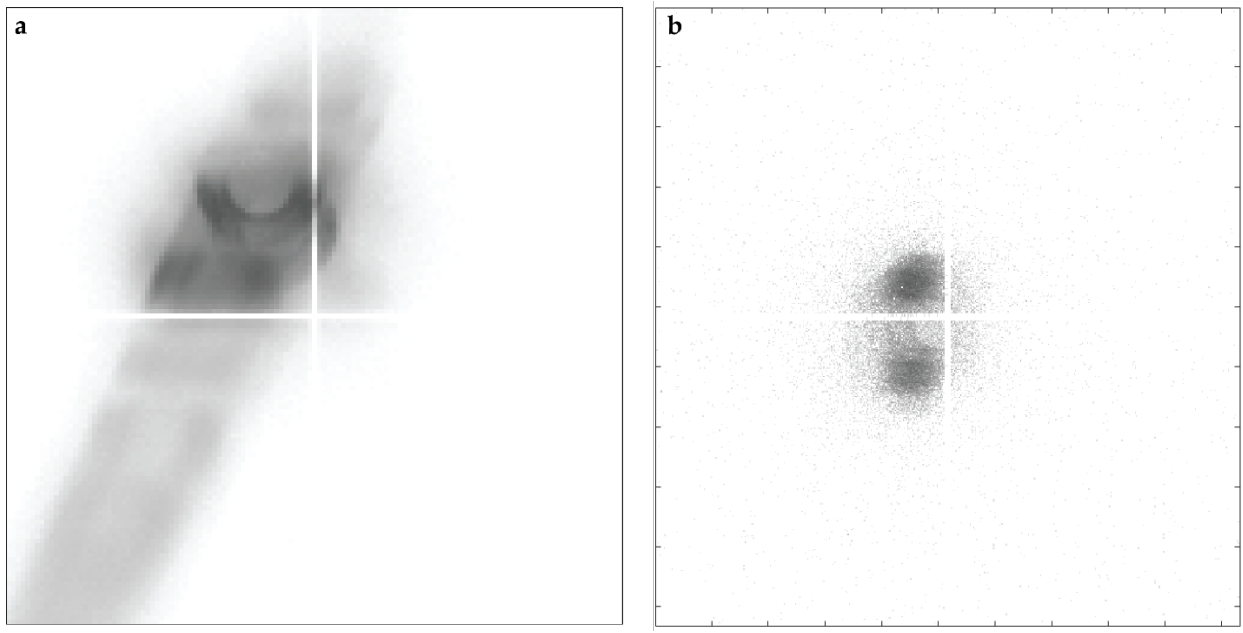

**Fig. S7.**
Enlarged averaged nanodiffraction patterns at 250 K (A) and 300 K (B).

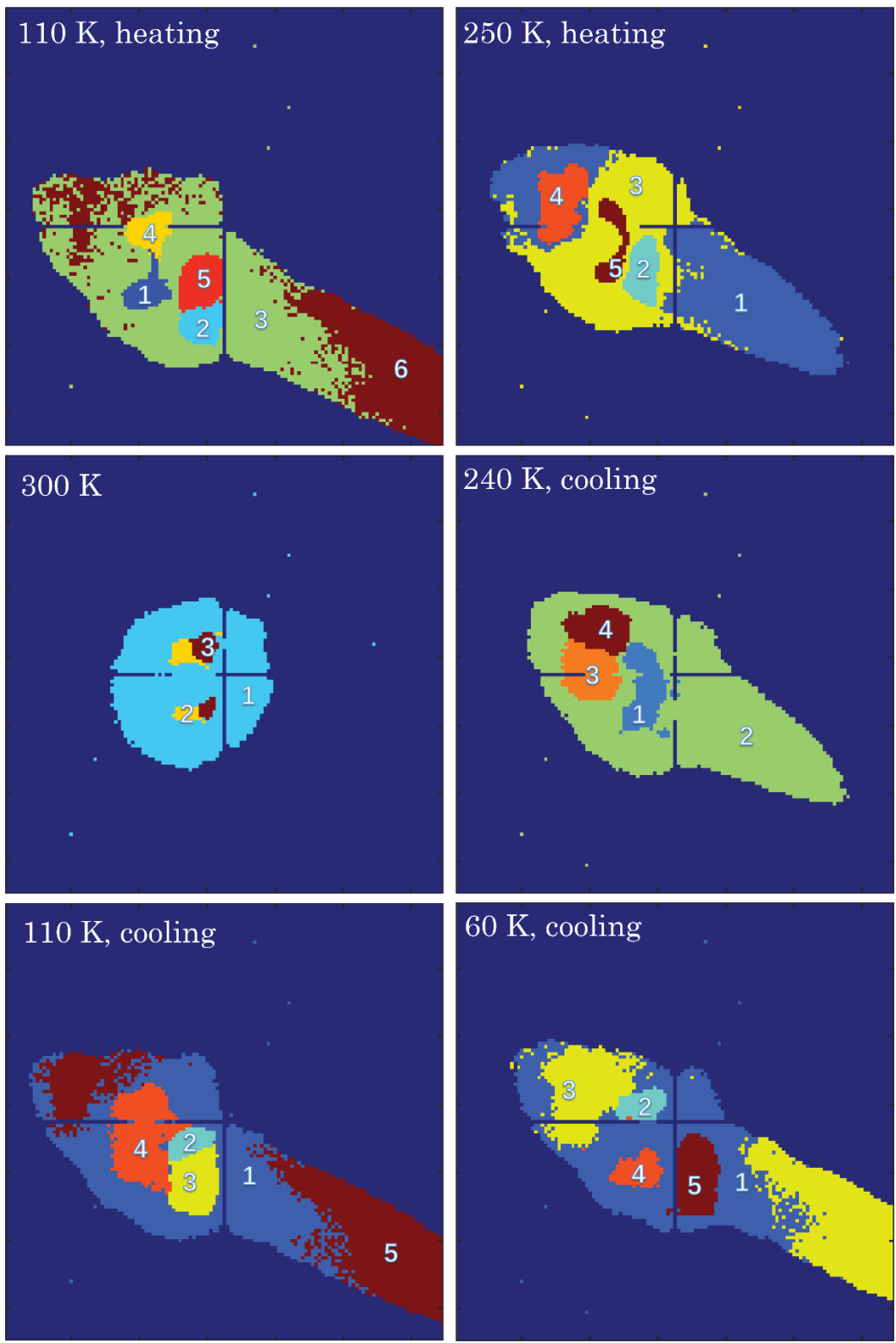

**Fig. S8.**
Clustering results at different temperatures through the MIT. Note a failure to separate low temperature phase scattering from background at 240 K, cooling.

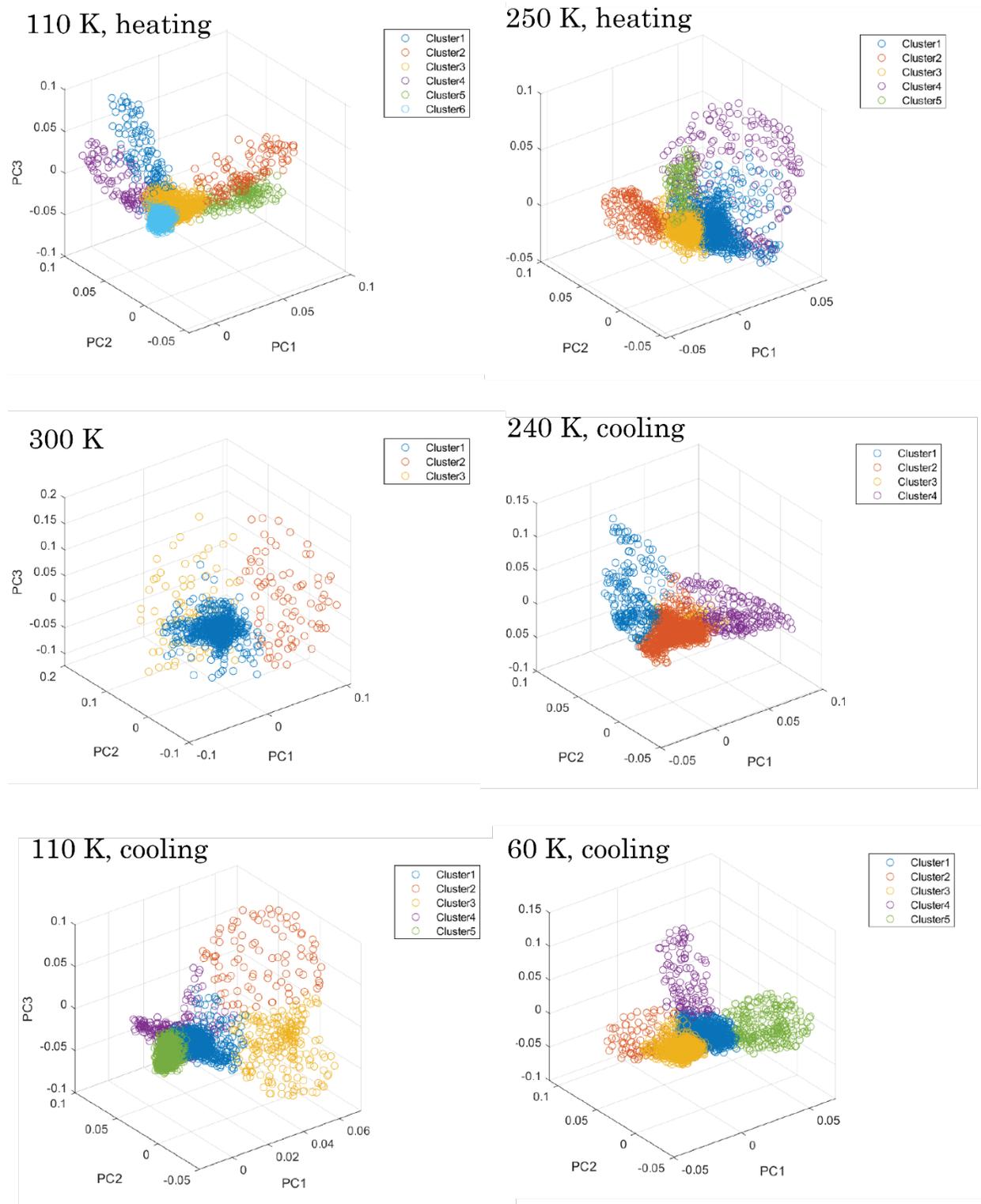

**Fig. S9.**
Clustering results at different temperatures through the MIT marked within 3 first principal components.